%Paper: astro-ph/9303005
%From: ats@het.brown.edu (Andrew Sornborger)
%Date: Wed, 10 Mar 93 11:11:05 EST

\input phyzzx
\hfuzz=5.0pt
\font\seventeenib=cmmib10 scaled\magstep3   \skewchar\seventeenib='177
\font\fourteenib=cmmib10  scaled\magstep2   \skewchar\fourteenib='177
\font\twelveib=cmmib10    scaled\magstep1   \skewchar\twelveib='177

\mathchardef\sperp="223F

\def\frac#1#2{\textstyle{#1\over #2}}

\def\.{.\thinspace}
\def\journal#1&#2(#3){\unskip,\enspace {\sl #1} {\bf #2} {\rm (#3)}}
\def\nextjournal#1(#2){{\bf #1} {\rm (#2)}}
\def\book#1#2(#3){{\sl #1\/} (#2, #3)}

%
%  Journal titles
%

\def\PL{Phys. Lett.}
\def\PR{Phys. Rev.}

\def\PRL{Phys. Rev. Lett.}

%
%  Publishers
%
%
%  References
%

\def\Borrill{J\.Borrill, E\.J\.Copeland and A\.R\.Liddle
        \journal\PL&B258 (1991) 310.}
\def\BorrillB{J\.Borrill, E\.J\.Copeland and A\.R\.Liddle
        {\it Phys. Rev. D} in press.}
\def\Cen{R\.Y\.Cen, J\.P\.Ostriker, D\.N\.Spergel, and N\. Turok, {\it
        Ap\.J\.}}

\def\Leese{R\.A\.Leese and T\.Prokopec
        \journal\PL&B260(1991) 27.}

\def\NumRec{{\it Numerical Recipes, the Art of Scientific Computing,}
        W\.H\.Press, B\.P\.Flannery, S\.A\.Teukolsky and W\.T\.Vetterling,
        Cambridge University Press}

\def\Perivolaropoulos {L\.Perivolaropoulos \journal\PR&D46(1992) 1858.}

\def\ProkopecB{T\.Prokopec, A\.Sornborger and R\.H\.Brandenberger
        \journal\PR&D45(1991) 1971.}

\def\TurokA{N\.Turok\journal\PRL&63(1989) 2625.}
\def\TurokB{N\.Turok and D\.Spergel\journal\PRL&64(1990) 2736.}
%
%   Start of paper
%
\FRONTPAGE\papers
\line{\hfill BROWN-HET-895}
\line{\hfill March 1993}

\title{A Semi-Analytical Study of Texture Collapse}

\author{A\.Sornborger}

\address{Department of Physics,\break
Brown University,\break
Providence,\break
RI 02912, USA.}

\abstract

This study presents a simplified approach to studying the dynamics of
global texture collapse. We derive equations of motion for a
spherically symmetric field configuration using a two parameter
ansatz. Then we analyse the effective potential for the resulting
theory to understand possible trajectories of the field configuration
in the parameter space of the ansatz. Numerical results are given for
critical winding and collapse time in spatially flat non-expanding,
and flat expanding universes. In addition, the open non-expanding and
open-expanding cases are studied.

\sequentialequations

\chapter{\bf Introduction}

\FIG\figansatz{The ansatz for $\chi$ as a function of $r$}

\FIG\figsaddle{A contour plot of the effective potential.}

\FIG\figflatrone{Trajectories are projected on $(\alpha, \xi)$-space.
In this figure trajectories are plotted for the spatially flat
non-expanding universe with $R=1.5$, $x=1.0$. All trajectories start on
the line $\xi=\alpha$}

\FIG\figexpandhalfrone{Trajectories are plotted for the
radiation-dominated universe with $R=1.5$ and $x=1.0$.}

\FIG\figexpandtwothirdsrone{Trajectories are plotted for the
matter-dominated universe with $R=1.5$ and $x=1.0$.}

\FIG\figflatrtwo{Trajectories are plotted for the flat case with
$R=2.0$ and $x=1.0$.}

\FIG\figcoltimeflat{Collapse time, defined as the time it takes for
$x$ to reach half its initial value, is plotted versus winding,
for a non-expanding universe.}

\REF\rTurokA{\TurokA}
\REF\rTurokB{\TurokB}
\REF\rLeese{\Leese}
\REF\rBorrillA{\Borrill}
\REF\rBorrillB{\BorrillB}
\REF\rProkopecB{\ProkopecB}
\REF\rCen{\Cen}
\REF\rPerivolaropoulos{\Perivolaropoulos}
\REF\rNumRec{\NumRec}

Global texture theory is a cosmological model of large-scale structure
formation. Texture is a semi-topological defect in a theory where a
global symmetry group $G$ is broken to a group $H$ such that
$\pi_3({G\over{H}})\neq 1$. We can take as a toy model a theory with a
complex doublet of scalar fields and a Mexican hat potential
generalized to four dimensions, in which case the vacuum manifold is
$S^3$.

Texture is formed when the field wraps around enough of $S^3$ to cause
unwinding. Energy lumping of the field configuration causes accretion
of matter, thus forming large-scale structure \refmark\rTurokA
\refmark\rTurokB.

To understand cosmological structure formation, we need to have
information about the texture distribution. The texture distribution
depends crucially on the critical winding \refmark\rLeese (i.e. the
winding above which collapse occurs). The formation probability for
texture with a given winding distribution depends on the critical
winding. More or fewer textures are formed if the critical winding is
low or high, respectively.

Critical winding is determined by the dynamics of the field
configuration as uncorrelated regions of the field come into the
horizon. Thus, we need to have a good understanding of field dynamics
in the texture model to understand the texture distribution.

Dynamics of texture field configurations has been analyzed in a
variety of studies. A self-similar $\sigma$-model solution has been
found for the flat, non-expanding universe with winding $w=1$
\refmark\rTurokA. For this solution collapse proceeds at the speed of
light. This study is limited by the fact that dynamics at the point of
unwinding are not obtainable with the approximations used. Also,
solutions for non-integer winding are not known.

Numerical studies have been able to get past the above limitations
\refmark\rBorrillA \refmark\rBorrillB \refmark\rProkopecB.
The full equations have been integrated for spherically symmetric and
ellipsoidal configurations in flat, expanding universes; non-integer
winding and the dynamics at the point of collapse were also included
in these studies. Furthermore, full cosmological simulations have also
been done, including baryonic matter effects \refmark\rCen.

Some analytical work has been done in the adiabatic approximation
using the $\sigma$-model approach \refmark\rPerivolaropoulos. These
studies have approximated the Hamiltonian for large $R$, where $R$ is
a physical cutoff understood to be half the intertexture separation,
and used a two parameter ansatz for a spherically symmetric field
configuration.

This study extends the above analytical work to include kinetic
energy. Here we investigate a constrained Hamiltonian, where the field
configuration is (as in the above study) approximated by a two
parameter ansatz.

First, we investigate the shape of the effective potential due to
gradient energy to understand what sort of trajectories can be
expected. Then, we integrate the Hamiltonian equations of motion to
find the trajectory followed by the field ansatz in phase space.

Analysis of the effective potential for this theory results in a clear
understanding of why trajectories in phase space evolve as they do. A
saddlepoint is found lying on the barrier between expanding and
collapsing trajectories explaining the existence of trajectories which
appear initially to be headed for collapse, but then expand, and
trajectories which appear to be headed toward expansion, yet finally
collapse (see section 3).

Critical windings were found for fields uncorrelated on scales larger
than the horizon for intertexture separation $2R$, where $R$ is in
units of the horizon size, to be:

For $R=2.0$ (For $R=1.5$) (errors are plus or minus the difference to
the next closest calculated trajectory.)

\centerline{$0.6562 \pm 0.0001$ ($0.6967 \pm 0.0001$) (flat case),}
\centerline{$0.6693 \pm 0.0001$ ($0.7471 \pm 0.0001$) (radiation era),}
\centerline{$0.667 \pm 0.007$ ($0.7543 \pm 0.0001$) (matter era).}

Collapse time was found to increase at critical winding when
approached from high winding. This phenomenon is also explained by the
saddlepoint found in the analysis of the effective potential. In fact,
critical winding should, according to arguments based on the shape of
the effective potential, go to infinity at critical winding.

\chapter{\bf Derivation of Equations of Motion}

As was mentioned in section 1, global texture can be described by a
theory with a complex doublet of scalar fields with a Mexican hat
potential. The action for this theory is

$$S=\int d^4x[\partial^\mu\Phi ^a\partial_\mu \Phi^a-\lambda
(|\Phi|^2-\eta^2)]\sqrt{-g}$$
where $a=1,...,4$ and $\eta$ is the scale of symmetry breaking.

For low temperatures relative to the phase transition and times before
unwinding we can fix the scalar field at the minimum of the potential
and treat the potential as a constraint. This gives us a
$\sigma$-model action
$$S=\int d^4x [\partial^\mu \Phi^a \partial_\mu \Phi^a]\sqrt{-g}$$
with the constraint
$$|\Phi|^2=\eta^2.$$

We can use the spherically symmetric ansatz
$$\Phi^a=\eta(\sin\chi\sin\theta\sin\phi,\sin\chi\sin\theta\cos\phi,
\sin\chi\cos\theta, \cos\chi),$$
where $\chi=\chi(r,t)$ is a radial field and the background Friedmann-
Robertson-Walker metric
$$g_{\mu\nu}=diag(1,-{a^2\over{1-kr^2}},-a^2r^2,-a^2r^2\sin^2\theta),$$
to obtain the action
$$S=\int d^4x[\dot\chi^2-{\chi'^2(1-kr^2)\over{a^2}}-
{2\sin^2\chi\over{a^2r^2}}]{a^3r^2\over{\sqrt{1-kr^2}}}.$$

In the flat ($k=0$) non-expanding case ($a(t)=1$), the solution to the
equations of motion is

\centerline{$\chi=2\arctan(-{r\over{t}})$ for $t<0$.}

We can mimic the exact solution with the simple ansatz (Figure
\figansatz)

\centerline{$\chi=\alpha\pi r$ for $0<r<x$}
\centerline{$\chi=\xi\pi$ for $x<r<R$}
and vary $\xi$ to obtain non-integer winding. In this ansatz
$\alpha=\alpha(t)$ and $\xi=\xi(t)$, $x$ is the point of intersection
of the two segments of the ansatz and $R$ is a cutoff. $x$ is
initially taken to be the horizon size (the scale on which the field
is correlated) and $R$ is taken to be $1/2$ the intertexture
separation, a fixed multiple of $H^{-1}$. The initial horizon $H(t_0)$
is set to be equal to $1$.

This ansatz is expected to be good for understanding precollapse
dynamics, especially around the time when the field configuration
enters the horizon.

With this ansatz in the above action, we derive the Hamiltonian
$$H=p_\alpha^2(4a^3\pi^2I_1)^{-1}+p_\xi^2(4a^3\pi^2I_4)^{-1}$$
$$+a\alpha^2\pi^2I_2+2aI_3+2a\sin^2(\xi\pi)I_5$$
where $p_\alpha$ and $p_\xi$ are momenta canonical to $\alpha$ and
$\xi$ respectively, and $I_i$ are integrals over the radial variable
$r$.

{}From this Hamiltonian we find the equations of motion in phase-space
($p_\alpha,p_\xi,\alpha,\xi$) to be

$$\dot\alpha=p_{\alpha}(2a^3\pi^2I_1)^{-1}$$
$$\dot\xi=p_{\xi}(2a^3\pi^2I_4)^{-1}$$
$$\dot p_{\alpha}=p_{\alpha}^2(4a^3\pi^2)^{-1}I_1^{-2}{\partial
I_1\over{\partial\alpha}}+p_{\xi}^2(4a^3\pi^2)^{-1}I_4^{-2} {\partial
I_4\over{\partial\alpha}}$$
$$-2a\alpha\pi^2I_2-a\alpha^2\pi^2{\partial I_2\over{\partial\alpha}}
-2a{\partial I_3\over{\partial\alpha}}$$
$$-2a\sin^2(\xi\pi){\partial I_5\over{\partial\alpha}}$$
$$\dot p_{\xi}=p_{\alpha}^2(4a^3\pi^2)^{-1}I_1^{-2} {\partial
I_1\over{\partial\xi}} +p_{\xi}^2(4a^3\pi^2)^{-1}I_4^{-2} {\partial
I_4\over{\partial\xi}}$$
$$-a\alpha^2\pi^2{\partial I_2\over{\partial\xi}}-2a{\partial
I_3\over{\partial\xi}}$$
$$-4\pi a\sin(\xi\pi)\cos(\xi\pi)I_5-2a\sin^2(\xi\pi){\partial
I_5\over{\partial\xi}},$$
and the integrals $I_i$ are
$$I_1=\int^x_0 dr {r^4\over{\sqrt{1-kr^2}}}$$
$$I_2=\int^x_0 dr r^2 \sqrt{1-kr^2}$$
$$I_3=\int^x_0 dr {\sin^2(\alpha\pi r)\over{\sqrt{1-kr^2}}}$$
$$I_4=\int^R_x dr {r^2\over{\sqrt{1-kr^2}}}.$$
$$I_5=\int^R_x {dr\over{\sqrt{1-kr^2}}}$$

The integrals are all trivial except $I_3$ in the open case ($k\neq
0$). For this case, we expand $\sin^2r$ to sufficient order and
calculate the resulting integrals. To solve the equations of motion in
the open case, we need expressions for $a(t)$, the scale factor of the
universe. We use the fact that the universe is close to flat up to its
current point of evolution and use the flat space radiation- and
matter-dominated expanding universe scale factors as approximations to
the actual scale factor.

Finally, the expression relating $\xi$ to winding $w$ is
$$w(\xi)=\xi-{\sin(2\pi\xi)\over{2\pi}}.$$

\chapter{\bf Results}

The terms in H independent of momenta $p_\alpha$ and $p_\xi$ form the
effective potential due to gradient energy in the field. Here (Figure
\figsaddle) we plot a portion of it for the flat non-expanding case
with $R=1.5$.

There are two attractors in the potential, one in the upper righthand
corner (large $\alpha$, large $\xi$) to which collapsing
configurations approach; and one in the lower lefthand corner (small
$\alpha$, small $\xi$) to which expanding configurations approach.

Notice the saddlepoint. To the lower left (small $\alpha$, small
$\xi$) and upper right (large $\alpha$, large $\xi$) are the
attractors for collapsing and expanding configurations. To the upper
left (small $\alpha$, large $\xi$) and lower right (large $\alpha$,
small $\xi$) we have regions of high potential.

If $R$ is increased, the saddlepoint moves leftward to small values of
$\alpha$ with $\xi$ remaining approximately the same, and its size
shrinks (that is the gradient gets steeper on all sides). If $R$ is
decreased, the saddlepoint moves to the right to larger values of
$\alpha$ with $\xi$ remaining approximately the same, and its size
increases (gradients get less steep on all sides). The gradients
around the saddlepoint get large for small $R$ because the total
energy of the configuration increases as $R$ gets large and thus the
height of the barrier increases.

We take initial conditions for $\xi$ and $\alpha$ based on the
physical constraint $x_{initial}=1$. This means that the initial $\xi$ and
$\alpha$ obey the relation $\xi=\alpha$. For intertexture separation
$2R=4$, this implies that for winding near critical the starting point
of the field evolution trajectory is on the upper left region (hill)
of the saddlepoint.

For our initial conditions, $x_{initial}=1$ and thus
$\dot\alpha\sim p_\alpha$ and $\dot\xi\sim p_\xi$, so near the
saddlepoint, the field momenta can be thought of as `rolling' on the
effective potential (farther away momentum behavior becomes more
complex).

Therefore the field will roll down the hill and up the barrier on the
opposite side of the saddle. Depending on whether the initial
configuration is above or below critical winding the trajectory will
scatter off the upper or lower side of the righthand barrier. If the
trajectory moves to the basin in the upper right of the potential, the
configuration collapses and unwinds. But if the trajectory scatters
off the lower part of the righthand barrier, the configuration
collapses for some time, but then, as the configuration falls to the
lower left attractor, the configuration expands.

For large $R$, the opposite effect occurs. Initially trajectories
start on the lower right hill, then they roll across toward the
opposite barrier and scatter to one side or the other depending on
initial winding. These trajectories were discovered by
Perivolaropoulos \refmark\rPerivolaropoulos in the large $R$ limit.

Trajectories that don't start on the saddlepoint fall more or less
directly into the attractors on their respective sides of the
saddlepoint.

For large $R$, there are fewer trajectories that start on or near the
saddlepoint because its size shrinks.

Monte Carlo simulations suggest that \refmark\rLeese, if we assume
random correlations on length scales larger than the horizon,
intertexture separation should be around two times the horizon size.
This indicates that, usually, few textures will expand then collapse,
but many will first tend towards collapse then expand, due to the
location of the saddlepoint with respect to typical initial
conditions.

The following figures show integrated trajectories with
$x_{initial}=1$ for flat non-expanding, flat expanding and open
expanding universes. The trajectories are plotted in
($\xi,\alpha$)-space.

Critical winding can be found either by distinguishing which
trajectories finally expand and which finally collapse, or by
calculating the intersection of the line of initial conditions and the
line of intersection of the tangent surface to the top of the ridge in
which the saddlepoint lies with the ridge. We found the former method
more convenient.

Note that critical winding increases from the non-expanding
(Figure \figflatrone) to the radiation-dominated (Figure
\figexpandhalfrone) to the matter-dominated cases (Figure
\figexpandtwothirdsrone). This can be understood physically. The
expanding background introduces an extra pull on the texture
configuration, thus causing more configurations to expand, and thus
pushing critical winding higher than in the non-expanding universe.
The matter-dominated background expands more quickly than the
radiation-dominated background, thus pushing critical winding even
higher in the matter-dominated case.

The effect of extra pull due to the background manifests itself in the
effective potential such that the barrier region, including the
saddlepoint moves to a higher value of $\xi$.

In an open background, the area of concentric spheres increases more
quickly as we leave the origin than in a flat background. This effect
also tends to add a pull to the configuration, and push critical
winding upwards.

As $R$ increases (Figure \figflatrtwo), the saddlepoint in the
effective potential moves to the left in the figures. This causes the
line of initial conditions to move across the saddlepoint and critical
winding (the point of intersection of the line of initial conditions
and the top of the barrier) goes down, since the saddlepoint is in a
skewed orientation with the lefthand region of high potential at
larger $\xi$ than the righthand region of high potential.

Critical windings were found for fields uncorrelated on scales larger
than the horizon ($x_{initial}=1$) for intertexture separation $2R$,
where $R$ is in units of horizon size, to be:

For $R=2.0$ (for $R=1.5$) (errors are plus or minus the difference to
the next closest calculated trajectory.)

\centerline{$0.6562 \pm 0.0001$ ($0.6967 \pm 0.0001$) (flat case),}
\centerline{$0.6693 \pm 0.0001$ ($0.7471 \pm 0.0001$) (radiation era),}
\centerline{$0.667 \pm 0.007$ ($0.7543 \pm 0.0001$) (matter era).}

Collapse time is affected by the length of time the configuration
takes to roll off the saddle. Configurations very near to the
saddlepoint take longer to roll off, since the saddlepoint is a point
of unstable equilibrium. For configurations starting at the unstable
equilibrium point, collapse time should be infinite.

Also, there are trajectories below critical winding that evolve for a
substantial period of time as if they would collapse, but eventually
expand. Although there is no definable collapse time for these
configurations, there might be appreciable matter accretion possible.

Collapse time, defined as the time it takes for $x$, the joining point
of the two segments in the ansatz, to reach half its initial value,
for the flat non-expanding universe is plotted in Figure
\figcoltimeflat.

\chapter{\bf Discussion and Concluding Remarks}

We have used a two parameter spherically symmetric ansatz in the low
temperature $\sigma$-model approximation to the texture action to
obtain a Hamiltonian and equations of motion for texture dynamics.

{}From the Hamiltonian we have isolated the effective potential and
found a saddlepoint that explained the behavior of the field
configuration. The saddlepoint explains configurations which expand
then collapse, and configurations which collapse then expand, as well
as simpler trajectories in which the field configuration simply
collapses or simply expands.

{}From the Hamiltonian we derived equations of motion which were
ordinary differential equations, in contrast to partial differential
equations. The ODEs were easy to integrate on the computer, and
results were obtained $\sim 1000$ times faster than integrating the
PDEs.

Critical windings were found for fields uncorrelated on scales larger
than the horizon ($x_{initial}=1$) for intertexture separation $2R$,
where $R$ is in units of horizon size, to be:

For $R=2.0$ (For $R=1.5$) (errors are plus or minus the difference to
the next closest calculated trajectory.)

\centerline{$0.6562 \pm 0.0001$ ($0.6967 \pm 0.0001$) (flat case),}
\centerline{$0.6693 \pm 0.0001$ ($0.7471 \pm 0.0001$) (radiation era),}
\centerline{$0.667 \pm 0.007$ ($0.7543 \pm 0.0001$) (matter era).}

Collapse time was found to increase as winding approached critical
from above (for collapsing configurations).

This study is also interesting, in general, as an approach to studying
the $\sigma$-model for non-integer windings where there is a cutoff
scale $R$.

\ack
I am grateful to Robert Brandenberger, Robert Leese, Leandros
Peri- volaropoulos and Tomislav Prokopec for important conversations.
This work has been supported in part by a NASA fellowship and in part
by DOE grant DE-AC02-03130 Task A.

\endpage
\refout
\endpage
\figout
\bye